# Influence of pulsed laser deposition growth conditions on the thermoelectric properties of $Ca_3Co_4O_9$ thin films

Hank W. Eng[1], W. Prellier[1]*, S. Hébert[1], D. Grebille[1], L. Méchin[2], and B. Mercey[1]

[1]Laboratoire CRISMAT, CNRS UMR 6508, ENSICAEN,

6 Boulevard du Maréchal Juin, F-14050 Caen cedex, FRANCE

[2]Laboratoire GREYC, CNRS UMR 6072, ENSICAEN and the University of Caen,

6 Boulevard du Maréchal Juin, F-14050 Caen cedex, FRANCE

**Abstract**

Thin films of the misfit cobaltite $Ca_3Co_4O_9$ were grown on (0001)-oriented (*c*-cut) sapphire substrates, using pulsed-laser deposition techniques. The dependence of the thermoelectric/transport properties on the film growth conditions was investigated. The optimal conditions (for low resistivities) were found to be 600 °C, 0.1 – 0.2 mbar of oxygen pressure, and 1.7 J/cm$^2$. These films exhibited slightly metallic behaviour, consistent with in-plane resistivity curves of single crystals and *c*-axis magnetically aligned samples. Hall effect measurements showed the density of the hole-like carriers was $5.5 \times 10^{19}$/cm$^3$. The in-plane epitaxial relationship between the thin film and the sapphire substrate are investigated.

* prellier@ensicaen.fr

# I. INTRODUCTION

Thermoelectric materials convert heat into electricity (and vice versa) through Peltier and Seebeck effects and can be used for refrigeration, cooling microelectronic devices, and providing an energy source from waste heat.[1,2] Current thermoelectric materials are unstable at high temperatures. For the practical purpose of generating electricity from waste heat, these thermoelectric materials should perform well between the temperatures of 600 and 1000 K (Ref. 3). The quality of thermoelectric materials is quantified by the unitless figure of merit (*ZT*) which is calculated via Equation 1, where *T*, *S*, *r*, and *k* are temperature, thermoelectric power (or Seebeck coefficient), resistivity, and thermal conductivity, respectively. Thus, good thermoelectric materials should have high Seebeck coefficients, low resistivities, and low thermal conductivity.

$$ZT = \frac{S^2}{rk} \qquad (1)$$

Conventional thermoelectric materials today include metal chalcogenides (such as $Bi_2Te_3$), transition metal disilicates, and silicon-germanium alloys, which have figures of merit between 0.4 and 1.3 (Ref. 1). At high temperatures these compounds are unstable due to oxidization, decomposition, and/or melting. To economically compete against compressors and environmentally unfriendly chlorofluorocarons, these materials need a figure of merit of ~3 (Ref. 3).

Although far more stable than the aforementioned compounds, oxides were not thoroughly investigated in the past because conventional thermoelectric wisdom predicted these compounds would have low figures of merit due to their low carrier densities (as is expected from their ionic structures). The recent discovery of relatively high thermoelectric power (~100 *m*V/K at 300 K) and low resistivity (0.2 mΩ.cm) in an oxide ($NaCo_2O_4$) by Terasaki *et al.*[4] has ushered in a promising new area of thermoelectric oxide research[3,5,6,7,8,9]. The misfit cobaltite $Ca_3Co_4O_9$ has more potential than $NaCo_2O_4$ due to its high thermoelectric power (~125 *m*V/K at 300 K) and stability in air[7].

$Ca_3Co_4O_9$ can be denoted as $[Ca_2CoO_3]^{RS}[CoO_2]_{1.62}$ to recognize the two layers that constitute the misfit structure and the ratio of the differing *b*-axis parameters. The two layers are a $Ca_2CoO_3$ rock salt-like layer (RS) and a $CoO_2$ cadmium iodide-like layer (see Fig. 1). These two layers have similar *a* and *c* lattice parameters with differing *b* lattice parameters, where the ratio of the *b* parameters for the $Ca_2CoO_3$ layer to $CoO_2$ layer is 1.62. The $Ca_2CoO_3$ layer is a distorted rock salt layer with in-plane Co-O distances of ca. 2.28 Å and out-of-plane Co-O distances of 1.82 and 1.89 Å. The $CoO_2$ layer has edge-sharing $CoO_6$ octahedra with Co-O distances of ca. 1.86 and 1.96 Å[7,10,11,12].

Thin films offer the prospects of enhancing thermoelectric property of a material via quantum confinement effects, reduction of lattice thermoconductivity using lattice mismatches, and/or thermionic emissions between layered heterostructures[13], as well as improvement of thermopower based upon structural modifications. Quantum wells would require very thin layers (on the order of nanometer thickness) to reduce thermal conductivity, and thus, significantly improve thermoelectric properties.[14] Electronic band structures for $NaCo_2O_4$ reveal the bands near the Fermi energy are sensitive to distortions in the $CoO_6$ octahedra[15], and modification of the $CoO_2$ layers has been suggested by Maignan *et al.*[16] as the source of the improved thermoelectric properties of $[Pb_{0.4}Ca_{2.0}Co_{0.6}O_3]^{RS}[CoO_2]_{1.61}$ when compared to $[Pb_{0.7}Sr_{2.0}Co_{0.3}O_3]^{RS}[CoO_2]_{1.8}$. More specifically, with this misfit compound, strain between the cobaltite and the substrate may slightly alter the parent geometry of the $CoO_2$ layers to produce better thermoelectric properties. Recently Zhou *et al.* made $(Ca_{2.6}Bi_{0.4})Co_4O_9$ films on MgO substrates using a spin coating technique and found that smaller grain sizes could be correlated with higher thermopower, larger resistivity, and lower carrier concentrations[17]. In this article, we report the preparation and characterization of thin films of the misfit cobaltite $Ca_3Co_4O_9$ on *(0001)*-oriented sapphire. These films were prepared to investigate (1) how thermoelectric properties of the films would compare with previous bulk polycrystalline and single crystal studies and (2) how those properties are correlated to growth conditions and grain sizes of the films.

## II. EXPERIMENTAL MEASUREMENTS

A black, sintered $Ca_3Co_4O_9$ sample, synthesized using conventional solid state synthesis techniques[7], was used as the target for the pulsed-laser deposition. *(0001)*-oriented (*c*-cut) sapphire substrates obtained from Crystec were cleaned in acetone and ethanol, heated for twenty minutes in ca. 75 °C solution of 3:1 $H_2SO_4/H_3PO_4$, and then finally washed in water. The acidic solution was made from Carlo Erba 96 wt% sulfuric acid and Acros 85 wt% phosphoric acid solutions. Sapphire and magnesium oxide were both initially used as substrates, but depositions onto MgO were problematic due to an impurity phase with *2?* diffraction peaks at 37.5° and 42.22°. For this reason we chose to deposit films only on sapphire, which is a good candidate for two reasons: both have trigonal or close-to-trigonal symmetries (in the $CoO_2$ layer) and the lattice mismatch is relativelly small. This lattice mismatch is easily seen in the next-nearest oxygen distances in the rock salt-like $Ca_2CoO_3$ layer (4.558 Å), $CoO_2$ layer (4.834 Å) and sapphire (4.761 Å)[11,18]; therefore, a theoretical mismatch of the bulk to sapphire distances would be at least 1.5% with the $CoO_2$ layer and 4.5% with the $Ca_2CoO_3$ layer. The deposition chamber was first evacuated to a base pressure on the order of $10^{-8}$ mbar, and then the substrate was heated to 600 °C. Oxygen pressure was subsequently increased to ca. 1 mbar, and the substrate was heated for thirty minutes more in this atmosphere to anneal the cleaned surface. Next, the chamber was brought to the deposition temperature and oxygen pressure. A KrF excimer laser beam (Lambda Physik compex, $\lambda$ = 248 nm, repetition rate = 3 Hz) was used to deposit the target onto the sapphire substrate between temperatures of 550 and 650 °C, energies of 1.4 and 1.9 $J/cm^2$, and oxygen pressures of 0.1 and 0.3 mbar. Finally after the deposition, the oxygen pressure was increased again to ca. 1 mbar before the chamber was cooled.

Thin film depositions were characterized by X-ray diffraction (XRD), and the films' transport properties and thermoelectric powers were measured using a Quantum Design Physical Property Measurement System (PPMS). A Seifert XRD 3000P diffractometer was used to collect diffraction data over the two-theta range of 5 to 75° (0.01° step size, 0.5 s/step) using monochromatic Cu $K\alpha_1$ radiation ($\lambda$ = 1.54056 Å). The texture and orientations of the films were characterized by X-ray diffraction using pole figure measurements. These were obtained with a four circle X'Pert Philips Material Research Diffractometer using Cu $K\alpha$ radiation. A $DekTak^3$ ST surface profiler was used to measure the thickness of the samples, which were found to be ~1800 Å. Electrical resistivities were measured using the standard four-probe method at a current of 1 mA, from 10 to 400 K. To make the resistivity measurements, four plots of silver were first deposited onto the films by thermal evaporation, and then thin aluminum contact wires were subsequently bonded to the silver plots on the films to the electrodes. Finally, for the Hall effect measurements, a silver layer and then a gold layer were deposited onto the surface of the film before standard ultraviolet photolithography and argon ion etching were used to pattern microbridges (where the largest widths were 100 $\mu$m), which were then bonded using aluminum contact wires. The silver layer was deposited via thermal evaporation, and the gold layer was deposited via the rf sputtering technique.

## III. RESULTS

Figure 2 shows the $\theta - 2\theta$ scan of a representative thin film on the sapphire substrate, and the inset shows the $\omega$-scan rocking curve recorded around the *(002)* diffraction peak. From the seven *(00l)* peaks of the film, the *c*-axis parameter is calculated to be 10.71(2) Å, relatively close to the bulk value (10.76 Å)[10,11]. This smaller lattice parameter and the relatively large full-width-at-half-maximum (FWHM) of the rocking curve (0.83°) is likely due to the lattice mismatch between the monoclinic cobaltite and the rhombohedral/trigonal sapphire substrate.

In order to characterize the orientation and the crystalline quality of the film, pole figures of specific diffraction reflections were measured. The $(10\bar{1}4)$ reflection (Fig. 3a) of the substrate gives the orientation of the $a^*$-axis and equivalent directions of the trigonal structure of the sapphire. Of the thin film, two different reflections, $(\bar{1}12)$ and *(203)*, have been chosen for the first rock-salt like sublattice and the $(\bar{1}14)$ reflection has been chosen for the second $CoO_2$ sublattice. The three corresponding poles figures are shown in Fig. 3. Fig. 3c and 3d are specific for each sublattice of the structure and confirm the same misfit character for the film as for the bulk samples. The pole figures clearly show a good crystalline quality of the film with rather punctual poles (? angle FWHM is approximately 4°). Although the symmetry of the figure is hexagonal, the structure is monoclinic, so the resulting figure can only be explained by 6 twinned domains related to the trigonal symmetry of the substrate, rotated at 60° from each other. Comparing Fig.3a and 3c or 3d, it appears that the common *a*-direction of the film is aligned with the *{120}* directions of the substrate. In Fig. 3b, the thin poles denoted S1 and S2 are related to "pollution" of the $(\bar{1}12)$ pole figure of the film by the $(11\bar{2}3)$ and $(01\bar{1}2)$ intense lines of the substrate (*i.e.*, both the substrate and film have almost the same *?* and *f* values at these points) and confirm the previous epitaxy relations. The second sublattice corresponds to the $CoO_2$ layers of the structure and is characterized by a pseudohexagonal symmetry which is observed with the $(\bar{1}14)$ pole figure in Fig. 3d. This last reflection is characterized by a *2q* value very close to the *(203)* reflection (common to both sublattices), but can be clearly distinguished by a different *y* value in the pole figure. In summary, the *[100]* direction of the film aligns epitaxially with the six equivalent *<120>* directions of the sapphire substrate.

Thin films of $Ca_3Co_4O_9$ could be deposited onto sapphire at various deposition conditions: temperatures of 550 to 650 °C, fluences of 1.4 and 1.9 J/cm$^2$, and oxygen pressures of 0.1 to 0.3 mbar. Because the figure of merit of thermoelectric power (*ZT*) is inversely proportional to resistivity (see Equation 1), resistivity values were used to determine the optimal growth conditions of the thin films. As seen in Fig. 4, the minimal resistivities of the films deposited at 600 °C can be found with a deposition energy of approximately 1.7 J/cm². The oxygen pressure during deposition appears to influence the variability of the film resistivities, although the actual minimum resistivity does not significantly vary with the oxygen pressure.

The inset of Fig. 4 shows the correlation between grain sizes and film resistivities. From the Scherrer formula (Equation 2), the grain size *(D)* can be calculated from the Bragg angle *($\theta_B$)*, full-width-at-half-maximum of the Bragg angle *($FWHM_B$)*, and wavelength of the radiation *($\lambda$)*.

$$D = \frac{0.9\lambda}{FWHM_B \cos\theta_B} \quad (2)$$

The lowest resistivities were found for grain sizes between 60 and 80 nm (see inset of Fig. 4). This result contrasts with Zhou *et al.*'s study on the grain size dependence of bismuth-doped misfit cobaltite films on MgO substrates[17]. In that study, the smallest resistivities were correlated to the largest grain sizes, although it should be noted that their study was based upon using only temperature to affect the material's properties, whereas, our study also takes into account oxygen pressure and laser deposition energies. From hereafter, the data presented is for the films made at the optimal conditions: 600 °C, 0.1 – 0.2 mbar of oxygen, and 1.7 J/cm².

Transport properties of the thin film at room temperature are similar to the bulk properties. The films of calcium cobaltite on sapphire exhibit slightly metallic behaviour above 180 K. This metallic behaviour has been observed in both the in-plane resistivity measurements of the single crystals[7] and the magnetically *c*-axis-oriented bulk[19,20]. The resistivity at 300 K is 21.8 mΩ.cm, comparable to the bulk $Ca_3Co_4O_9$ (~12 mΩ.cm) but higher than in the magnetically aligned samples (~7 mΩ.cm). The minimum resistivity was found to be 21.4 m$\Omega$.cm at 180 K, higher than both the bulk (~7 m$\Omega$.cm, 70 K) and the magnetically aligned samples (~3 m$\Omega$.cm, 50 K).

Without increasing oxygen pressure, some films exhibited similar metallic behaviour, while others exhibited the semiconductor behaviour of the out-of-plane resistivities. As commonly observed in perovskite thin films, increasing the oxygen pressure is necessary during the cooling period to obtain the correct oxygen stoichiometry, which results in metal-like behaviour. Similarly, only metallic curves were found when the oxygen pressure was increased after making the deposition. The higher oxygen pressure likely decreases the amount of oxygen vacancies in the resulting film, and more oxygenated bulk samples of $Ca_3Co_4O_9$ have been found to have lower resistivities (and slightly higher themoelectric power).[21] Subsequent measurements of the same films revealed slightly higher resistivities than prior measurements, presumably due to the loss of excess oxygen from the thin film. All the transport properties reported herein were measured within two weeks of the deposition of the films.

Thermoelectric power measurements were measured by a steady-state method. The film was heated on one side to create a temperature gradient, and two separated differential thermocouples were used to measure both the temperature gradient and voltage drop (see inset of Fig. 5). Because temperature and voltage are measured at each point on the film, errors arising from heating of the sapphire substrate were minimized. Thermoelectric power as a function of temperature is shown in Fig. 5. Because the film resistivity increases dramatically below 100 K, the resistivity becomes too high for an accurate measurement of the thermoelectric power with our instrument at low temperatures. As seen in the bulk, the positive value indicates $p$-type-like carriers. The thermoelectric power at 300 K (110 $\mu$V/K) is smaller than the bulk (125 $\mu$V/K). Perhaps creating a layering of this misfit cobaltite and another material will allow thermionic emissions between layers, allowing more efficient cooling of the film, and thus, increase thermoelectric properties[22].

Finally, Hall effect measurements were performed (see Fig. 6). At 300 K, the Hall effect coefficient and hole-like carrier density were calculated to be $9.1\times10^{-9}$ m$^3$/C and $6.8\times10^{20}$/cm$^3$, respectively, in agreement with an estimated carrier density for [Bi$_{0.87}$SrO$_2$]$_2$[CoO$_2$]$_{1.82}$ (Ref. 23). Assuming one electron from each Co-site in the CoO$_2$ and Ca$_2$CoO$_3$ layers is a carrier in the crystal structure[11], the theoretical carrier density from each layer can calculated to be $1.36\times10^{22}$/cm$^3$ and $8.46\times10^{21}$/cm$^3$, respectively. Even though the misfit cobaltite shows metallic behaviour, Ca$_3$Co$_4$O$_9$ is still a semiconductor, so the number of carriers is expected to be much lower than the number of available carriers.

## IV. CONCLUSIONS

In summary, thin films of the misfit cobaltite Ca$_3$Co$_4$O$_9$ were grown on (0001)-oriented sapphire substrates using pulsed-laser deposition techniques. To determine which films were the best candidates for thermoelectric power, we correlated the film growth conditions to resistivity properties. For the range of temperatures (550 – 650 °C), energies (1.4 – 1.9 J/cm$^2$) and oxygen pressures (0.1 – 0.3 mbar) that Ca$_3$Co$_4$O$_9$ grew on sapphire, we found the minimum resistivity occurred at 600 °C and 1.7 J/cm$^2$, regardless of deposition pressure of oxygen. For the optimized film, the room temperature resistivity was 21.4 m$\Omega$.cm and exhibited slight metallic behaviour, consistent with in-plane resistivity curves of single crystals and *c*-axis magnetically aligned samples. Higher resistivities were found for subsequent measurements, presumably due to oxygen diffusing out of the films. Although the thermoelectric power measurements were relatively close to the bulk value, theoretically this property may be improved by creating a superlattice between this misfit cobaltite and another material.  This artificial superlattice and its resulting properties will be investigated in future endeavors based on this $Ca_3Co_4O_9$ film.

## VI. ACKNOWLEDGEMENTS

We would like to thank Drs. A. Maignan and Ch. Simon for helpful discussions. One author (HWE) would like to further thank Dr. P. Padhan and S. Autier-Laurent for technical aide.  Funding for this project was provided by the Centre National de la Recherche Scientifique (CNRS) and Conseil Régional Basse Normandie.

Figure Captions

Figure 1. $Ca_3Co_4O_9$ structure. This misfit cobaltite is also known as $[Ca_2CoO_3]^{RS}[CoO_2]_{1.62}$ because of the $Ca_2CoO_3$ and $CoO_2$ monoclinic layers which have similar *a*, *c*, and **b** lattice parameters but differing *b* lattice parameters. The ratio of the *b* parameters is 1.62.

Figure 2. X-ray diffraction results of the $Ca_3Co_4O_9$ thin film on (0001)-oriented sapphire substrate. The *c*-axis was calculated to be 10.71(4) Å, approximately 1% shorter than in bulk studies[7,11]. The relatively large FWHM of the rocking curve (inset) may be due to the 1.5% lattice mismatch between the monoclinic misfit cobaltite to hexagonal sapphire substrate (see text).

Figure 3. X-ray pole figures: (a) $(10\bar{1}4)$ pole of the sapphire substrate; (b) $(\bar{1}12)$ and $(1\bar{1}2)$ poles of the first sublattice of the misfit structure of the film (F1 and F2). S1 and S2 are related to the *(113)* and *(012)* poles of the substrate; (c) *(203)* pole of the first rock salt-like sublattice; (d) $(\bar{1}14)$ pole of the second $CoO_2$ sublattice.

Figure 4. Room temperature resistivity dependence upon grain size (inset), energy, and millibars of oxygen pressure at 600 °C for $Ca_3Co_4O_9$ thin films on sapphire. Regardless of the oxygen pressure, the minimal resistivities occur at energies of approximately 1.7 $J/cm^2$ with grain sizes between 60 and 80 nm. The lines are drawn to help guide the reader's eyes for isobaric depositions in the main figure and resistivity-to-grain size trends in the inset.

Figure 5. Resistivity and Seebeck coefficient versus temperature measurements for the Ca$_3$Co$_4$O$_9$ thin film on sapphire. Resisitivity measurements show metallic behaviour in the range from 180 to 300 K. The arrows indicate the resistivity measurement as a progression of temperature. Seebeck measurements of the film could not be performed below 50 K because the resistivity increases dramatically at that temperature. The inset shows the setup for the thermoelectric measurement.

Figure 6. Hall effect measurements of the Ca$_3$Co$_4$O$_9$ thin film on sapphire: resistivity versus magnetic field for various temperatures. The positive slope indicates the p-type-like behaviour of the carriers.

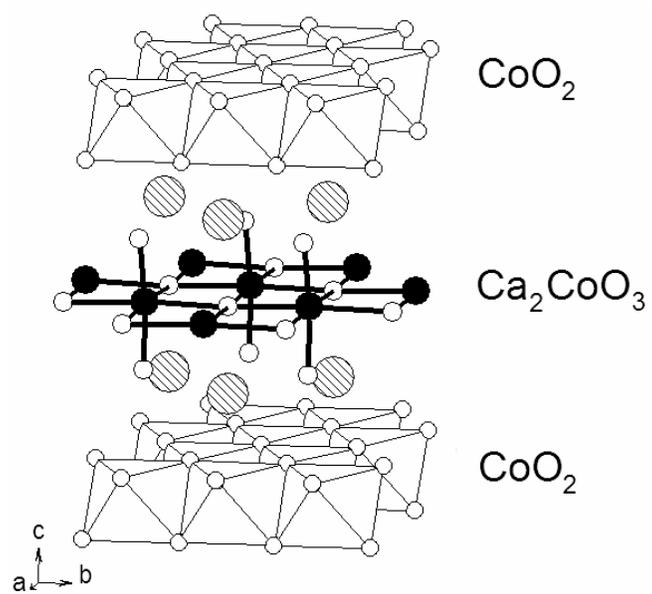

Figure 1

Eng *et al*.

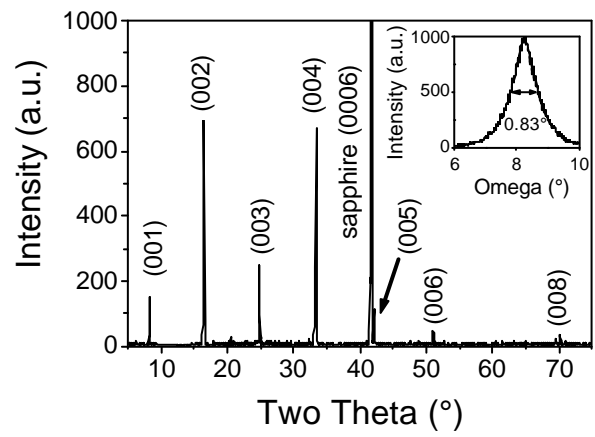

Figure 2

Eng *et al.*

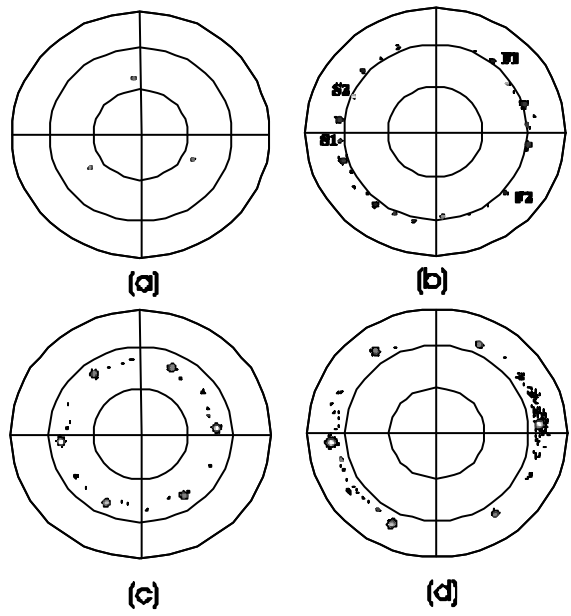

Figure 3

Eng *et al.*

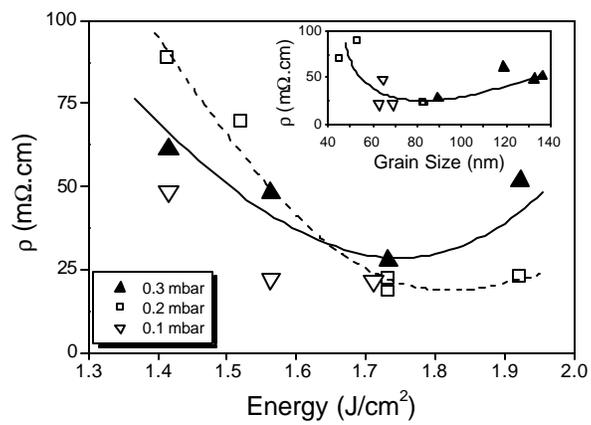

Figure 4

Eng *et al.*

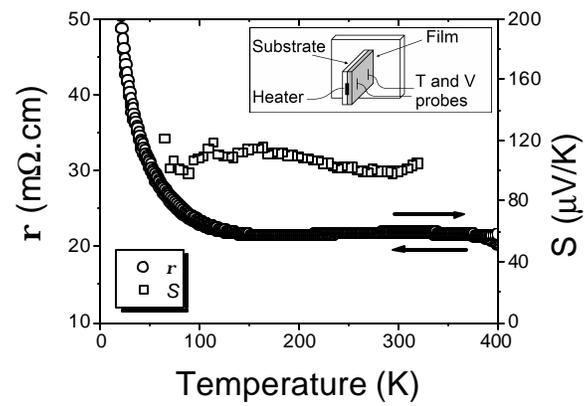

Figure 5

Eng *et al.*

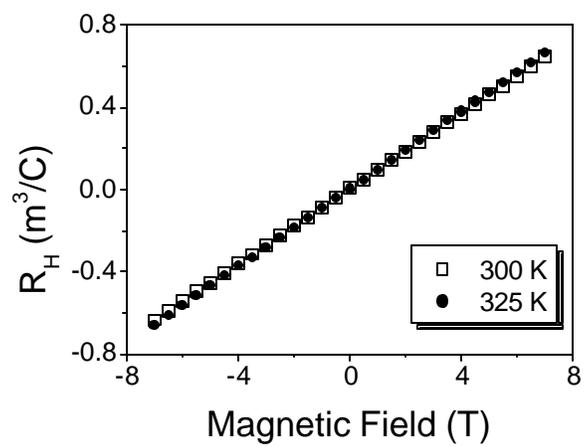

Figure 6

Eng *et al.*